\documentclass[prl,superscriptaddress,twocolumn]{revtex4}

\pdfoutput=1

\usepackage{amsmath}
\usepackage{amsfonts}
\usepackage{amssymb}
\usepackage{graphicx}
\usepackage{color}

\begin{document}

\title{Three Dimensional Broadband Tunable Terahertz Metamaterials}
\date{\today}\received{}
\author{Kebin Fan}
\affiliation{Department of Mechanical Engineering, Boston University, 110 Cummington Street, Boston, Massachusetts 02215, USA}
\author{Andrew~C.~Strikwerda}
\affiliation{Department of Physics, Boston University, 590 Commonwealth Avenue, Boston, Massachusetts, 02215, USA}
\author{Xin Zhang}
\email{xinz@bu.edu} \affiliation{Department of Mechanical Engineering, Boston University, 110 Cummington Street, Boston, Massachusetts 02215, USA}
\author{Richard~D.~Averitt}
\email{raveritt@bu.edu} \affiliation{Department of Physics, Boston University, 590 Commonwealth Avenue, Boston, Massachusetts, 02215, USA}

\begin{abstract}
We present optically tunable magnetic 3D metamaterials at terahertz (THz)
frequencies which exhibit a tuning range of $\sim$30$\%$ of the resonance
frequency. This is accomplished by fabricating 3D array structures consisting of double-split-ring resonators
(DSRRs) on silicon-on-sapphire, fabricated using multilayer electroplating. Photoexcitation
of free carriers in the silicon within the capacitive region of the DSRR
results in a red-shift of the resonant frequency from 1.74 THz to 1.16 THz.
The observed frequency shift leads to a transition from a magnetic-to-bianisotropic
response as verified through electromagnetic simulations and parameter retrieval. Our approach
extends dynamic metamaterial tuning to magnetic control, and may find applications
in switching and modulation, polarization control, or tunable perfect absorbers.

\end{abstract}

\maketitle

The advent of artificial
electromagnetic materials has provided unique routes to engineer photonic media, leading to
the realization of intriguing phenomena. This includes the successful demonstration of invisibility
cloaking \cite{schurig06, valentine09, ergin10} and negative refraction \cite{shelby01, valentine08}. Scale invariance of the
underlying equations enables translation of phenomena realized in one region of
the electromagnetic spectrum to others. Metamaterials and in particular, split-ring resonators (SRR),  can couple to the electric, magnetic or both fields of incident electromagnetic
radiation. This leads to a Lorentzian-like effective response in the permittivity, permeability, and bianisotropy, described by the general constitutive relations as:
\begin{equation}
\left(\begin{array}{c}
$\textbf{p}$ \\
$\textbf{m}$ \\ \end{array} \right) =
\left(\begin{array}{cc}
\alpha_{EE} \ \ \ \ \ \    \alpha_{EH} \\
\alpha_{HE} \ \ \ \ \ \  \alpha_{HH}\\ \end{array} \right)
\left(\begin{array}{c}
\textbf{\emph{E}$_{in}$} \\
\textbf{\emph{H}$_{in}$} \\ \end{array} \right)
\end{equation}
where $\alpha_{EE}$ ($\alpha_{HH}$) is the tensorial purely electric (magnetic) polarizability,
quantifying the electric (magnetic) dipole induced purely by a incident electric (magnetic)
field and the off-diagonal $\alpha_{EH}$ ($\alpha_{HE}$) is the tensorial magnetoelectric coupled polarizability,
quantifying the electric (magnetic) dipole induced by an incident magnetic (electric) field. The Onsager relations
require $\alpha_{EH} = - \alpha_{HE}^T$ \cite{marques02}. The bianisotropic parameter $\xi$ relates to the magnetoelectric
polarizability by $\xi = ic_0\alpha_{EH}$, where $c_0$ is the velocity of light in free space.

\begin{figure}[ptb]
\begin{center}
\includegraphics[width=3.4in,keepaspectratio=true]%
{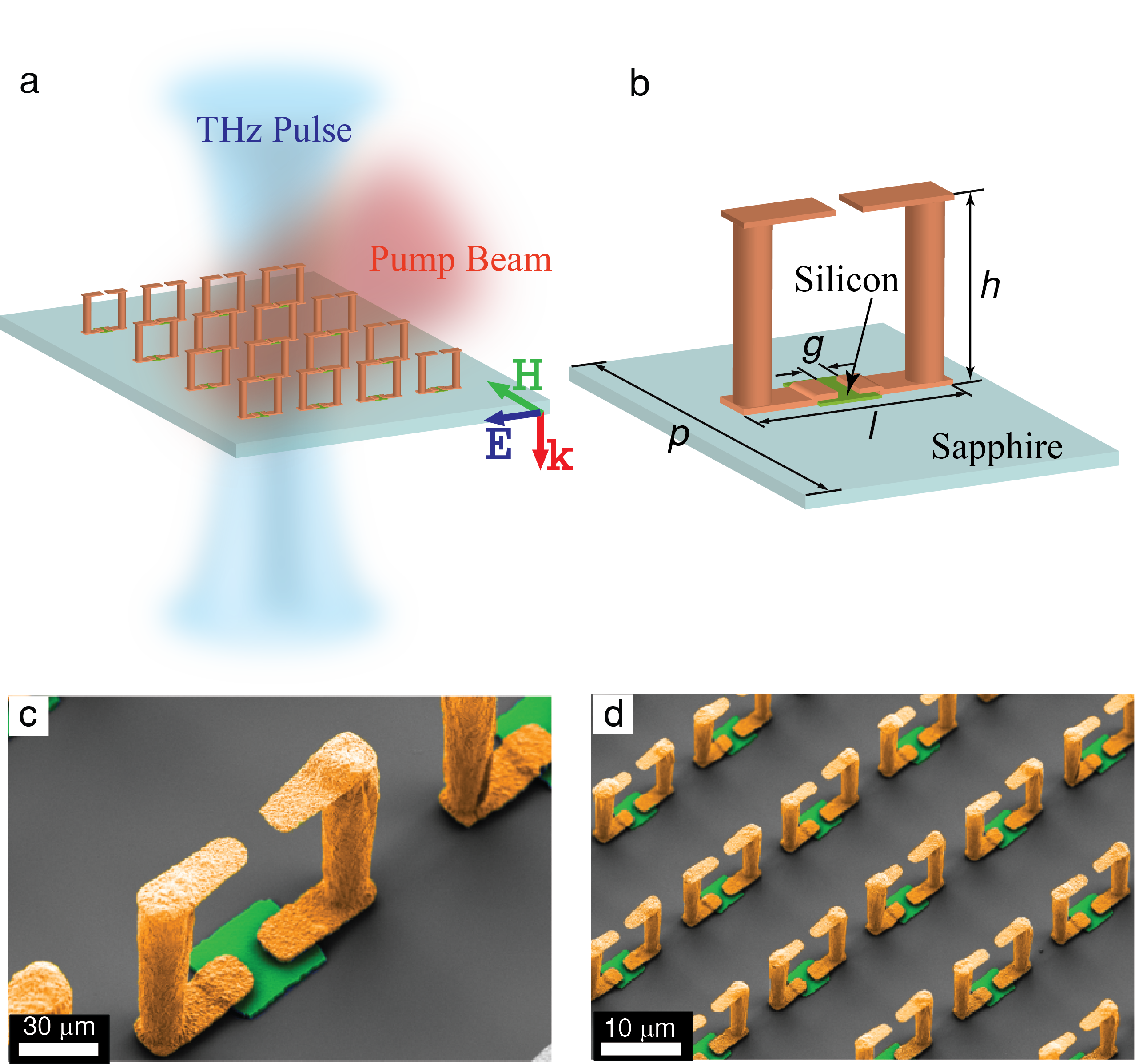}%
\caption{(Color online) Schematic of the design of the broadband tunable three dimensional
metamaterials at THz frequencies and scanning electron micrographs (SEM) of fabricated 3D
structures. (a) Schematic principle of the photoexcitation tuning of the 3D metamaterials. (b)
Typical dimensions for unit cell of DSRRs: p = 50 $\mu$m, h = 33 $\mu$m, g = 5$\mu$m, and l = 36 $\mu$m. (c)
The oblique close-up view of fabricated unit cell of DSRRs in copper standing on a sapphire substrate. (d) Oblique view of arrayed hybrid 3D metamaterials.}%
\label{Fig1}%
\end{center}
\end{figure}

By tailoring the geometry or configuration of subwavelength metallic unit cells, the effective parameters can be specified and controlled,
leading to myriad routes to control the amplitude, frequency,  and phase of the incident electromagnetic radiation.
For example, by independently tuning the electric and magnetic resonance
of multilayer composites, it is possible to match the impedance to free space
forming a perfect absorber with minimized reflectance \cite{landy08, tao08a, liux10}.
This ability to manipulate incident radiation is particularly
important at terahertz frequencies given the relative paucity of devices. Recent
progress on tunable metamaterial through voltage \cite{chen06, chen09}, optical
\cite{padilla06, chen08, shen11}, thermal \cite{driscoll08, chen10, zhu11}, and
mechanical \cite{tao09, zhuw11} tuning, has shown its potential to fill the ``THz gap".

The majority of previous studies are based on planar metamaterials, and therefore result in
tuning the electric response. It is desirable to explore new configurations to access magnetic control
of metamaterials which will, in turn, provide additional flexibility in tailoring the electromagnetic response for potential applications such
 as tunable absorbers or polarizers. Recently, three dimensional metamaterials at terahertz
frequencies have been demonstrated with the successful demonstration of negative
refraction \cite{zhang09} and negative permeability \cite{fan11},  with one example of a tunable three
dimensional (3D) metamaterial  \cite{zhangs12}. We demonstrate
 optically tunable magnetic 3D metamaterials exhibiting a tuning range of $\sim$30$\%$ of the resonance
frequency. This is accomplished by fabricating 3D array structures consisting of double-split-ring resonators
(DSRRs) on silicon-on-sapphire, fabricated using multilayer electroplating. Photoexcitation
of free carriers in the silicon within the capacitive region of the DSRR
results in a red-shift of the resonant frequency from 1.74 THz to 1.16 THz.  Further, the observed
frequency shift results in a transition from a magnetic-to-bianisotropic response as verified through electromagnetic simulations and parameter retrieval.

\begin{figure}[ptb]
\begin{center}
\includegraphics[width=2.5in,keepaspectratio=true]%
{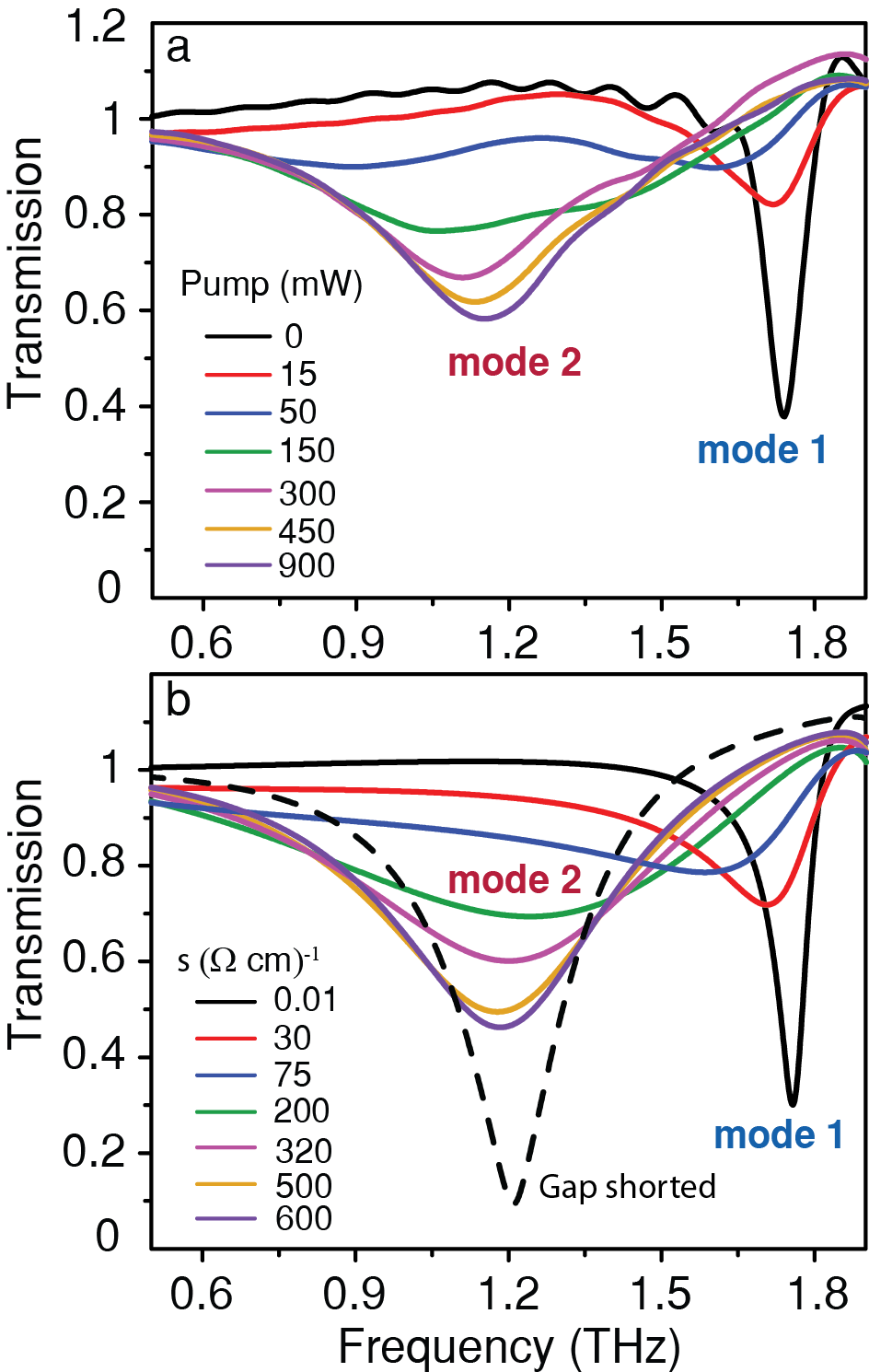}%
\caption{ (Color online) (a) Experimentally measured frequency dependent THz electric field transmission. The response is shown for increasing laser excitation power.
(b) Full wave electromagnetic simulation results of the transmission versus frequency for various values of the silicon conductivity. The dashed
black curve is a simulation with the bottom gap shorted with copper.}%
\label{Fig2}%
\end{center}
\end{figure}

The tuning principle and unit cell of the three dimensional metamaterial structure are depicted in
Fig. \ref{Fig1} (a) and (b). The arrayed 3D structures consist of stand-up DSRRs with gaps on the top and bottom, forming two capacitors
connected in series in an equivalent LC circuit. Silicon, the photoactive material in these structures,
is incorporated into the bottom gap of the DSRR. This structure is resonant when the THz wave is normally incident on
the sample with the E field parallel to the sides with the gaps  (H then passes through the DSRRs).
In the absence of photoexcitation, the THz pulses couple to the
resonators magnetically, inducing a circulating current in the ring. In particular, the symmetry
 is such that the electric field does not couple to the DSRRs (except for substrate-induced symmetry breaking as discussed below).
Upon optical excitation of the silicon, free carriers are generated resulting in an increased
conductivity. This decreases the capacitance of the lower gap which, in turn, increases the total
capacitance resulting in a redshift of the LC resonance. With
sufficient pump power, the bottom capacitor can be effectively shorted leading to
the maximum possible frequency shift. Importantly, with the lower gap shorted, the symmetry of the DSRR is reduced such that the incident
E field now drives circulating currents in the resonator (in addition to the H field). That is, photoexcitation induces intrinsic bianisotropy.

Multilayer electroplating (MLEP) in combination with conventional
optical lithography is utilized to fabricate the 3D metamaterials on a commercial
silicon-on-sapphire (SOS) wafer consisting of 600-nm (100) intrinsic silicon on top of a 530
$\mu$m-thick R-plane sapphire substrate. The fabrication of 3D DSRRs begins with
patterning of the distributed silicon pads using reactive ion etching (RIE). Then two
1.5-$\mu$m-thick bottom copper laminas were electroplated on a copper seed layer, forming
a 5-$\mu$m-wide gap between the two laminas. The laminas were patterned along the primary
flat orientation of the wafer, which is 45$^o$ counterclockwise from the projection of
the C-axis on the R-plane. Thus, the refractive indices on two axes are the same. The patterned
silicon pads sit beneath the gap. After removing the photoresist, a 30-$\mu$m
thick layer of photoresist (PR) AZ9260 was patterned with two 6-$\mu$m diameter holes
at both ends of the laminas to enable electroplating. Without removing the thick
PR, another 100-nm seed layer of copper was evaporated on the top without any wrinkling.
After one thin layer PR was patterned, 1.5-$\mu$m laminas that are the same as the bottom ones were
fabricated to connect the standing pillars. As a final releasing step, the PR was removed
with acetone, and the copper seed layers are etched away in dilute acid. Using MLEP,
 the height of the DSSRs can be determined by the thickness of PR,
which is controlled by the spinning rate. Fig. \ref{Fig1}(c) and (d) show the SEM images of the fabricated 3D metamaterials sample.

THz-TDS was used to characterize the frequency-dependent electromagnetic response
of the photoactive 3D-DSRRs. A bare sapphire substrate with the same thickness as the sample substrate
was used as the reference. The sample and reference were mounted on the holder with the
same orientation as to minimize sapphire birefringence. The THz pulses were at normal incidence to the substrate
with E-field parallel to the side of the DSRRs with the gaps (see Fig. 1(a)). Photoexcitation of free carriers in the silicon
was achieved with 1.55-eV, 35-fs ultrafast pulses at a repetition rate of 1 kHz. The optical pump pulse was set to arrive 10 ps before the THz probe beam
to achieve a near steady-state accumulation of carriers due to their long lifetime in silicon.
To ensure homogeneous excitation, the pump laser beam was 8 mm in diameter while the
incident THz had a 3 mm diameter.

The experimental results are shown in Fig. \ref{Fig2}(a). The black curve shows the response without
photoexcitation. There is a very sharp resonance at 1.74 THz ($\omega_{1}$) with a minimum
transmission of 35$\%$. As the pump power is increased to 450 mW, a pronounced resonance
shift to 1.16 THz ($\omega_2$) is observed with a transmission of 60$\%$. Further,
a tunability of 30$\%$ of the resonance frequency is achieved. At higher powers,
a large density of carriers result in significant carrier-carrier scattering, saturating the conductivity. For example, increasing the
excitation power from 450 mW to 900 mW) corresponding to a fluence of 1.8
mJ $cm^{-2}$) leads to only a minor change in the DSRR electromagnetic response.

Full wave electromagnetic  simulations were performed using CST microwave studio.
To compare with experiments, the conductivity of the silicon comprising the
substrate capacitor was varied in the simulations. The photoconductive silicon is simulated as dielectric with
$\epsilon_{si}$ = 11.7 and a pump-power dependent conductivity of $\sigma_{si}$. The
substrate sapphire was modeled as lossless dielectric with $\epsilon_{sapphire}$ = 10.5.
The copper was modeled as a lossy metal with a frequency independent conductivity of
5$\times$10$^{4}$ $(\Omega cm)^{-1}$, based on four-point probe measurements of
an electroplated continuous Cu film. The solid black curve in Fig. 3(b) corresponds to 0 mW pump
power with $\sigma_{si}$ taken as 0.01 $(\Omega cm)^{-1}$, while for illumination of 900 mW,
the conductivity is set as 600 $(\Omega cm)^{-1}$. As shown in Fig. \ref{Fig2}(b), the simulated and
measured frequencies match each other quite well, highlighting that the observed changes
in the electromagnetic response arise from changes in the Si conductivity.

\begin{figure}[ptb]
\begin{center}
\includegraphics[width=2.5in,keepaspectratio=true]%
{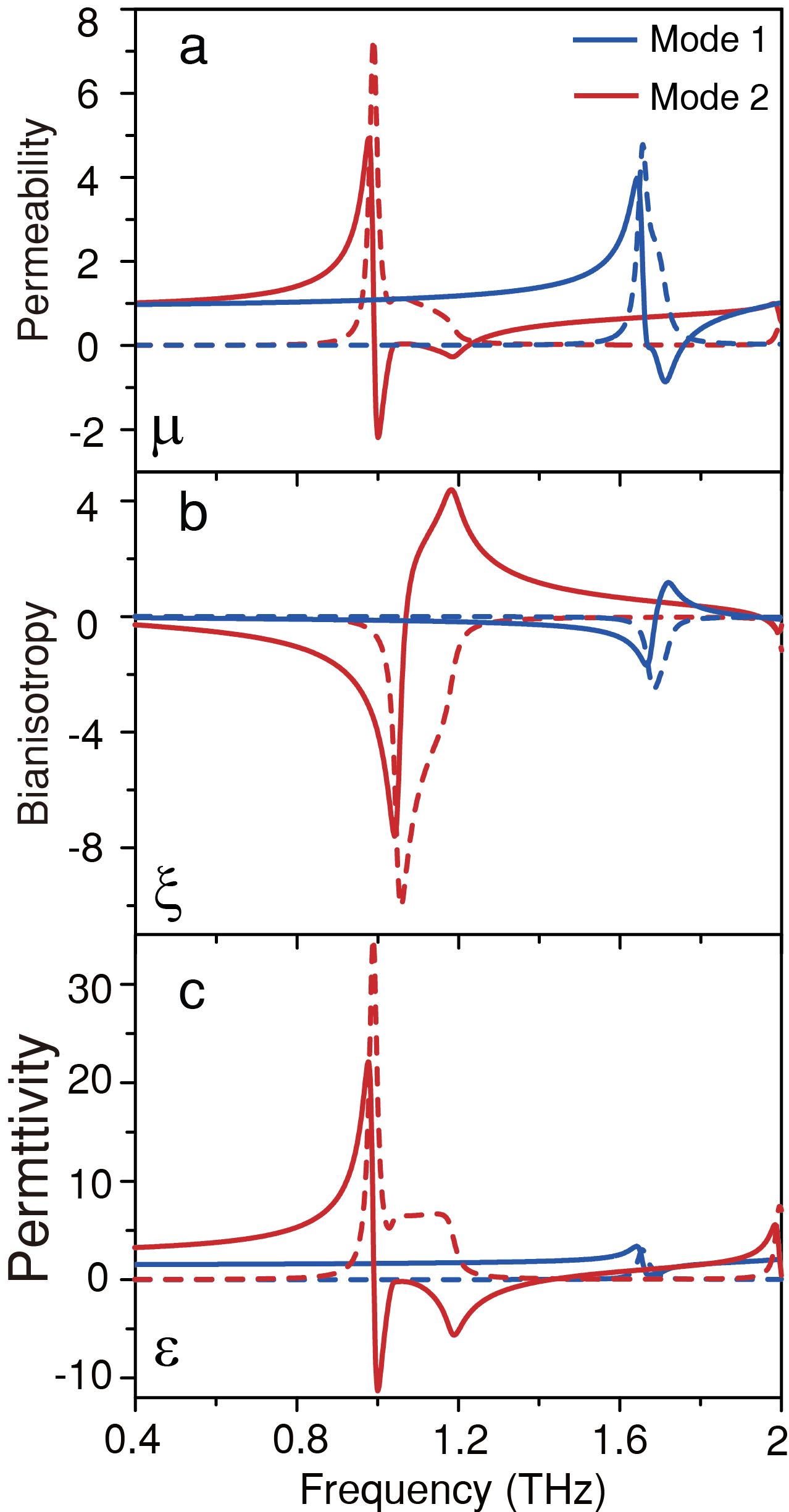}%
\caption{(Color online) Retrieved constitutive parameters of (a) permeability $\mu$, (b) bianisotropy
parameter $\xi$, and (c) permittivity $\epsilon$,  from full wave electromagnetic simulations of the 3D DSRRs.
The blue lines represent the parameters at mode 1 (i.e. no photoexcitation). The red lines represent the parameters at mode 2 assuming the bottom gap is shorted by
replacing the silicon by copper. Solid lines are real parts and dashed lines are imaginary parts.}%
\label{Fig3}%
\end{center}
\end{figure}

For our experiments, the electric field is normal to the gaps and the magnetic
field is normal to the plane of rings. As such, the electric and magnetic components of the THz waves
can, in principle, couple to the DSRRs leading to a bianisotropic response.
However, without photoexcitation (mode 1), inversion symmetry of the ring (in the propagation direction) eliminates the possibility
of electric field excitation resulting in a pure mangetic response (the substrate breaks this inversion symmetry
 and leads to measurable, but small, bianisotropy - see below and Fig. 3). Following photoexcitation, the shorted
bottom gap breaks the inversion symmetry, leading to strong electric field excitation and a
bianistropic response.

To further analyze the electromagnetic response of modes 1 and 2,
the real and imaginary parts of $\epsilon$, $\mu$ and $\xi$ were extracted from
simulations using a bianisotropic retrieval method \cite{kriegler09, zhao10, smith10}.  As
shown in Fig. \ref{Fig3}(b), at mode 1, the permeability exhibits a strong Lorentzian-like response at
the resonance yielding negative values in a certain span. However, as mentioned above, the
substrate breaks the symmetry and induces a small bianisotropy as the blue curves in Fig. \ref{Fig3}
(b) indicate \cite{powell10}. Hence, there is also a small resonant response for the permittivity
at mode 1. When the bottom gap is totally shorted (mode 2), the 3D DSRR array becomes fully
bianisotropic with both the electric and magnetic field resonantly driving the array.
The bianisotropic response is now dominated by the DSRRs. Although the permittivity and permeability present
negative values, the refractive index is still positive due to the strong bianisotropy.
Nonetheless, our metamaterials can be used to effectively switch the DSRRs with
 the electromagnetic response changing from near purely magnetic to bianisotropic.

\begin{figure}[ptb]
\begin{center}
\includegraphics[width=3in,keepaspectratio=true]%
{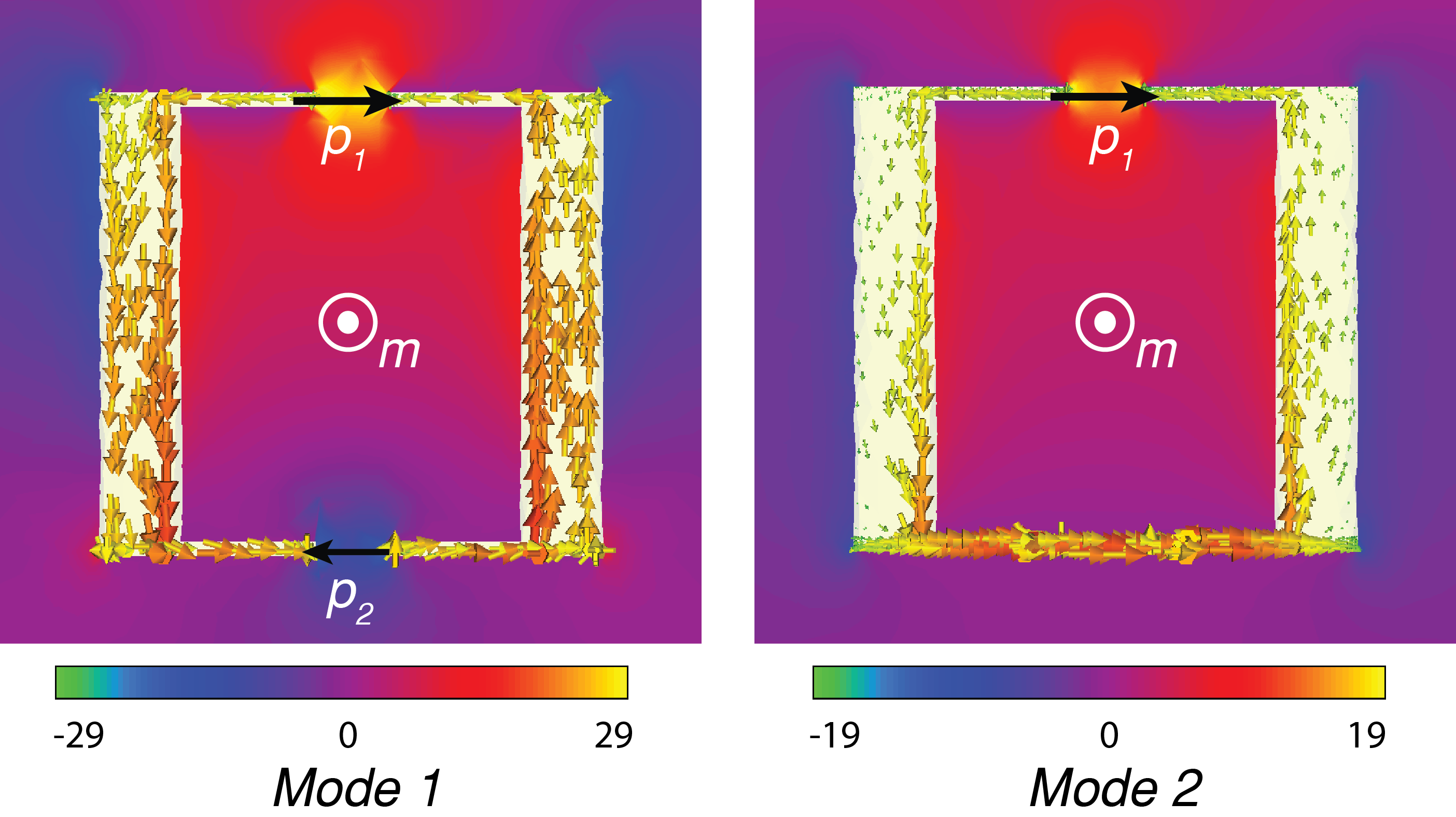}%
\caption{(Color online) Simulation of the distribution of surface current density and
electric field in the direction of normal to the gaps at two modes. Left: Surface
current and electric field at mode 1 with a magnetic dipole of $m$ and electric
dipoles of $p_1$ and $p_2$. Right: Surface current and electric field at mode 1
with a magnetic dipole of $m$ and an electric dipoles of $p_1$. The electric field and
the surface current are 90 degrees out of phase. The colorbars represent the simulated
excited field strength relative to the incident field.}%
\label{Fig4}%
\end{center}
\end{figure}

From the measurements and simulations, the mode 2 resonance is
broadened in comparison to  mode 1. While there are losses from excited carriers in the silicon, this is not the dominant source of the broadening. This is evident in Fig. 2(b) showing that the simulation  with the lower gap shorted with copper (black dashed line) is still considerably broader than the mode 1 resonance. In fact, the broadening of mode 2 results from increased radiation damping \cite{Li09}. Additional
 insight into the electromagnetic response and the nature of this damping is obtained through consideration of
the local electric field and current density (from simulations) at the unit cell level. Fig. \ref{Fig4}
shows the simulated surface currents and the electric field normal to the gap at both modes respectively. We note that the peak of
the surface currents is 90 degree out of phase with the peak of electric field. For mode 1, the circulating current causes charge accumulation at both gaps forming
two electric dipoles $p_1$ and $p_2$ in opposite directions, which can be equivalently
taken as a quadrupole as shown in the left side of Fig. 4. As such, radiation damping of the DSRRs for mode 1 is at the level of
magnetic dipole and electric quadrupole terms in a multipole expansion of the electrodynamic response.
 In contrast, for mode 2 (in which the bottom gap is effectively shorted), there is only
the top gap, which is equivalent to an electric dipole $p_1$ (right side, Fig. 4). Consequently, radiation damping for mode 2 is dominated by the
electric dipole contribution, leading to a decrease in the quality factor of the resonance. It is important to note that this is
intrinsic to the SRRs and is not a spurious contribution that can, for our geometry, be minimized.

In summary, we have successfully fabricated and characterized broadband tunable 3D
hybrid metamateirals at THz frequencies. With photoexcitation of silicon, we
observed over 30\% redshift of the resonant frequency, in agreement
with  numerical simulations. Parameters retrieval indicates that
our 3D structure can be effectively switched from a mode with a strong magnetic response
to a fully bianisotropic response. Our tunable 3D metamaterials can be potentially used
as a tunable notch filter, high speed modulator. With proper
design on the phase shift between two modes, artificial polarizers can also be realized.

We acknowledge partial support from DOD/Army Research Laboratory under
Contract No. W911NF-06-2-0040, NSF under Contract No. ECCS 0802036,
and AFOSR under Contract No. FA9550-09-1- 0708. The authors would like to thank the
Photonics Center at Boston University for all the technical support throughout the course of this research.


\begin{thebibliography} {99}

\bibitem{schurig06} D. Schurig \emph{et al.},  Science \textbf{314}, 977 (2006).

\bibitem{valentine09} J. Valentine, J. Li, T. Zentgraf, G. Bartal, and X. Zhang, Nature Mater. \textbf{8}, 568 (2009).

\bibitem{ergin10} T. Ergin, N. Stenger, P. Brenner, J. B. Pendry, and M. Wegener, Science \textbf{328}, 337 (2010).

\bibitem{shelby01} R. A. Shelby, D. R. Smith, and S. Schultz, Science \textbf{292}, 77 (2001).

\bibitem{valentine08} J. Valentine \emph{et al.}, Nature \textbf{455}, 376 (2008).

\bibitem{marques02} R. Marqu\`{e}s, F. Medina, and R. Rafii-El-Idrissi, Phys. Rev. B, \textbf{65}, 144440, (2002).

\bibitem{landy08} N. I. Landy, S. Sajuyigbe, J. J. Mock, D. R. Smith, and W. J. Padilla, Phys. Rev. Lett. \textbf{100}, 207402 (2008).

\bibitem{tao08a} H. Tao, \emph{et al.}, Opt. Express \textbf{16}, 7181 (2008).

\bibitem{liux10} X. Liu, T. Starr, A. F. Starr, and W. J. Padilla, Phys. Rev. Lett. \textbf{104}, 207403 (2010).

\bibitem{chen06} H.-T. Chen, \emph{et al.}, Nature \textbf{444}, 597 (2006).

\bibitem{chen09} H.-T. Chen, \emph{et al.}, Nature Photon. \textbf{3}, 148 (2009).

\bibitem{padilla06} W. J. Padilla, A. J. Taylor, C. Highstrete, M. Lee, and R. D. Averitt, Phys. Rev. Lett. \textbf{96}, 107401 (2006).

\bibitem{chen08} H.-T. Chen, \emph{et al.}, Nature Photon. \textbf{2}, 295-298 (2008).

\bibitem{shen11} N. H. Shen, \emph{et al.}, Phys. Rev. Lett. \textbf{106}, 037403 (2011).

\bibitem{driscoll08} T. Driscoll, \emph{et al.}, Appl. Phys. Lett. \textbf{93}, 024101 (2008).

\bibitem{chen10} H.-T. Chen, \emph{et al.}, Phys. Rev. Lett. \textbf{105}, 247402 (2010).

\bibitem{zhu11} J. Zhu, \emph{et al.}, Opt. Commun. \textbf{284}, 3129, (2011).

\bibitem{tao09} H. Tao, A. C. Strikwerda, K. Fan, W. J. Padilla, X. Zhang, and R. D. Averitt, Phys. Rev. Lett. \textbf{103}, 147401 (2009).

\bibitem{zhuw11} W. M. Zhu, \emph{et al.}, Adv. Mater. \textbf{23}, 1792 (2011).

\bibitem{zhang09} S. Zhang, \emph{et al.}, Phys. Rev. Lett. \textbf{102}, 023901 (2009).

\bibitem{fan11} K. Fan, A. C. Strikwerda, H. Tao, X. Zhang and R. D. Averitt, Opt. Express \textbf{19}, 12619 (2011).

\bibitem{zhangs12} S. Zhang, \emph{et al.}, Nature Commmun. \textbf{3}, 942 (2012).

\bibitem{kriegler09} C. E. Kriegler, M. S. Rill, S. Linden, and M. Wegener, IEEE J. Sel. Top. Quantum Electron. \textbf{16}, 367 (2010).

\bibitem{zhao10} R. Zhao, T. Koschny, and C. M. Soukoulis, Opt. Express \textbf{18}, 14553 (2010).

\bibitem{smith10} D. R. Smith, Phys. Rev. E \textbf{81}, 036605 (2010).

\bibitem{powell10} D. A. Powell,  and Y. S. Kivshar, Appl. Phys. Lett. \textbf{97}, 091106 (2010).

\bibitem{Li09} T. Q. Li, \emph{et al.}, Phys. Rev. B \textbf{80}, 115113 (2009).


\end{thebibliography}
\end{document}